\documentclass[twocolumn 
]{svjour3}          
\smartqed  
\usepackage{url}

\usepackage{amsmath}
\usepackage{graphicx}
\usepackage{color}
\usepackage[colorlinks=true, pdfstartview=FitV, linkcolor=blue, citecolor=red, urlcolor=magenta, breaklinks]{hyperref}
\usepackage{mathabx}
\usepackage{soul} 
\setstcolor{red}
\usepackage{cancel} 

\usepackage{lineno}
\modulolinenumbers[3]

\renewcommand{\Vec}[1]{\mbox{\boldmath$#1$}}
\newcommand{\Ves}[1]{\mbox{\boldmath$_{#1}$}}
\journalname{Nonlinear Dynamics}

\begin{document}
\title{Orbital perturbation coupling of primary oblateness and solar radiation pressure\thanks{A preliminary version of this research was presented at the Third International Nonlinear Dynamics Conference---NODYCON 2023 held in Rome, Italy, 18--22 June 2023 \cite{LaraFantinoFlores2024}.}}
\titlerunning{Orbital perturbation coupling of primary oblateness and solar radiation pressure}  
\author{Martin Lara \and
        Elena Fantino \and
        Roberto Flores }
\authorrunning{M.~Lara \and E.~Fantino \and R.~Flores} 
\institute{M.~Lara \at
           Scientific Computation and Technical Innovation Center, University of La Rioja, 
           26006 Logro\~no, Spain \\
           Tel.: +34-941-299440 \\
           Fax: +34-941-299460 \\
           \email{mlara0@gmail.com}           
           \and
           E.~Fantino \at
           Aerospace Engineering Department, and Space Technology and Innovation Center, Khalifa University of Science and Technology, 
           P.O. Box 127788 Abu Dhabi, United Arab Emirates\\ 
           \email{elena.fantino@ku.ac.ae (corresponding author)}           
           \and
           R.~Flores \at
					 Aerospace Engineering Department, Khalifa University of Science and Technology, 
           P.O. Box 127788 Abu Dhabi, United Arab Emirates
}

\maketitle              

\begin{abstract}

Solar radiation pressure can have a substantial long-term effect on the orbits of high area-to-mass ratio spacecraft, such as solar sails. We present a study of the coupling between radiation pressure and the gravitational perturbation due to polar flattening. Removing the short-period terms via perturbation theory yields a time-dependent two-degree-of-freedom Hamiltonian, depending on one physical and one dynamical parameter. While the reduced model is non-integrable in general, assuming coplanar orbits (i.e., both Spacecraft and Sun on the equator) results in an integrable invariant manifold. We discuss the qualitative features of the coplanar dynamics, and find three regions of the parameters space characterized by different regimes of the reduced flow. For each regime, we identify the fixed points and their character. The fixed points represent frozen orbits, configurations for which the long-term perturbations cancel out to the order of the theory. They are advantageous from the point of view of station keeping, allowing the orbit to be maintained with minimal propellant consumption. We complement existing studies of the coplanar dynamics with a more rigorous treatment, deriving the generating function of the canonical transformation that underpins the use of averaged equations. Furthermore, we obtain an analytical expression for the bifurcation lines that separate the regions with different qualitative flow.

\end{abstract}

%
\section{Introduction}
The dynamics of natural and artificial bodies in the solar system is dominated by the Keplerian attraction of either the Sun or a different natural massive body. However, perturbations such as the non-sphericity of the primary body, tidal effects, or solar radiation pressure, may accumulate with time yielding notable changes with respect to Keplerian dynamics. For objects with high area-to-mass ratio, radiation pressure is an important effect \cite{Plummer1905,Robertson1937,Peale1966,BurnsLamySoter1979}. An example is the dust dynamics in planetary rings \cite{Lumme1972,PokornyDeutschKuchner2023}. Of special relevance for space technology are solar sails, a potential means of efficient spacecraft propulsion \cite{McInnes1999,HeiligersFernandezStohlmanWilkie2019}. 

The impact of radiation pressure has been acknowledged since the beginning of the space era. In particular, it was identified as the cause of disagreement between the predicted trajectory and the observed behavior of Vanguard I\footnote{\href{https://nssdc.gsfc.nasa.gov/nmc/spacecraft/display.action?id=1958-002B}{nssdc.gsfc.nasa.gov/nmc/spacecraft/1958-002B}; last accessed December 25, 2023} \cite{MusenBryantBailie1960} and, most notably, the Echo satellites\footnote{\href{https://space.jpl.nasa.gov/msl/QuickLooks/echoQL.html}{https://space.jpl.nasa.gov/msl/QuickLooks/echoQL}; last accessed December 25, 2023} \cite{ParkinsonJonesShapiro1960,ShapiroJones1960,JastrowBryant1960,ZadunaiskyShapiroJones1961}. This motivated theoretical research on its effects over the long-term motion of spacecraft \cite{Musen1960JGR,Kozai1961SAO,Cook1962,Aksnes1976}. In spite of the non-conservative character of radiation pressure, it can be approximated with a disturbing potential. Under this simplification, the problem may be approached with Hamiltonian dynamics, which is particularly useful in the context of resonant motion \cite{Musen1960JGR,Kaula1962,Brouwer1963IUTAM,FerrazMello1972,Hughes1977}. 

The long-term effects of radiation pressure in high area-to-mass objects have been described analytically with a surrogate integrable dynamics \cite{Mignard1982,MignardHenon1984,Deprit1984,MeerCushman1987,ChamberlainBishop1993}. The coupling with oblateness alters significantly the long-term dynamics \cite{Musen1960JGR,Brouwer1963IUTAM,Hamilton1993,HamiltonKrivov1996,KrivovGetino1997,FengHou2019}. This opens opportunities for the design of novel mission orbits \cite{ColomboLuckingMcInnes2012}, including deorbiting strategies \cite{GkoliasAlessiColombo2020,MasseSharfDeleflie2023}. 

The characterization of the coupled effect of the central-body oblateness and radiation pressure perturbations is, to our knowledge, still incomplete. Even in the simplest approach of the cannonball model \cite{Kubo1999} with constant solar flux and negligible solar parallax \cite{Kozai1961SAO}, which results in constant acceleration, the long-term behavior is governed by a time-dependent, two-degree-of-freedom system whose closed-form solution is not known. Notwithstanding the lack of a general solution, the dynamics of specific resonances have been discussed in detail \cite{AlessiColomboRossi2019,GkoliasAlessiColombo2020}. For the special case when the Sun lies on the equatorial plane, the coplanar orbits become an invariant manifold of the averaged problem. After truncation of higher-order effects, the reduced Hamiltonian depends on one physical and one dynamical parameter. Then, the general characteristics of the reduced flow can be studied in the plane of these parameters.

The standard analysis in literature starts directly from the averaged equations of orbital evolution. Then, the types of motion arising from different relative strengths of the governing parameters are studied. We establish a more formal framework for the problem with a complete canonical perturbation approach. We build the generating function of the infinitesimal contact transformation that removes the short-period terms from the original Hamiltonian \cite{Poincare1892vII,FerrazMello2007}. It provides the necessary theoretical foundation for the averaging assumptions \cite{Arnold1989}, and enables the computation of higher-order solutions \cite{Brouwer1959,Kozai1962,DepritRom1970,CoffeyDeprit1982,FerrerLara2010,LaraPerezLopez2017,Lara2020}. Beyond the qualitative description of the dynamics, the transformation between initial conditions and corresponding averaged variables is needed to initialize the constants of the perturbation theory. Their accurate computation is critical for the correct propagation of the long-term dynamics \cite{Cain1962,LyddaneCohen1962,Walter1967,BreakwellVagners1970,Lara2020arxiv}.

In the same spirit, seeking rigorous description of the typologies of motion, we derive analytical expressions for the fundamental lines of the parameters plane that separate regions with different types of flow. This gives a formal underpinning to the mechanisms controlling changes in the flow, both local---bifurcations of relative equilibria---and global---related to the evolution of orbits that eventually become circular---. We demonstrate a complete description of the reduced phase space in terms of arithmetic operations only: the fundamental lines are determined computing discriminants of polynomial equations and applying Descartes' rule of signs.
For each regime, we identify the fixed points and their character. These points represent frozen orbits, configurations where the long-term effects of radiation pressure and oblateness cancel out to the order of the theory. Therefore, while short-term perturbations (i.e., with a periodicity of one orbit) persist, the secular drift of the orbital elements vanishes. This has the potential to extend spacecraft operational life, allowing for long-term station keeping with minimal propellant consumption.

Another original contribution is the derivation of the averaged equations in vectorial form. It is free from singularities and enhances the stability of numerical integration \cite{LaraRosengrenFantino2020,SanJuanLopezLara2024}. Even though it introduces redundancy in the averaged differential system, computational cost does not increase because the symmetry of the equations allows for an efficient implementation.

The paper is organized as follows. After justifying the simplifications that yield the approximate Hamiltonian in \S\ref{s:pemo}, the perturbation solution is approached in \S\ref{s:Hamsol}, where the elimination of short-period terms provides a compact set of variation equations in vectorial form that can be efficiently integrated semi-analytically. Finally, the dynamics of the coplanar manifold are discussed in detail in \S\ref{s:coplanar}. The Wolfram Mathematica software provided assistance with mathematical manipulations and plotting of results.

\section{Perturbation model} \label{s:pemo}

\begin{figure}[htbp]
	\center
	\includegraphics[scale=0.3]{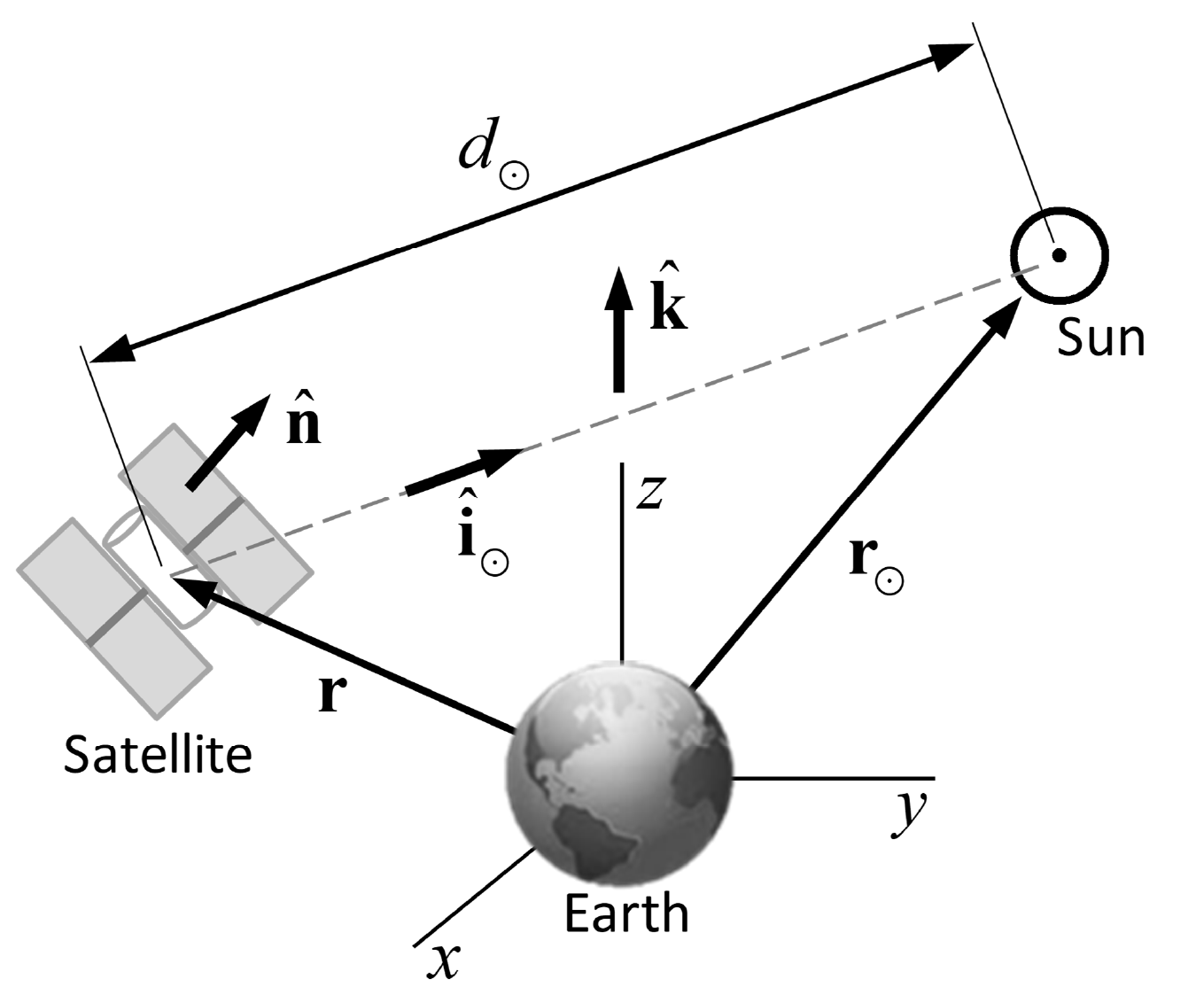}
	\caption{Vectors and distances used in the perturbation model.}
	\label{f:vectors}
\end{figure}

We focus on the case of a negligible mass object, the ``orbiter'', moving around an oblate central body, the ``planet'', at a distance where the tidal forces of the Sun, which we assume to revolve with Keplerian motion in its apparent orbit about the planet, are small compared to both radiation pressure and the effect of planetary oblateness. That is, $r/d_\Sun\ll1$, where
\begin{align}
r=& \; \sqrt{\Vec{r}\cdot\Vec{r}}, \\ \label{dsun}
d_\Sun=&\; \sqrt{(\Vec{r}_\Sun-\Vec{r})\cdot(\Vec{r}_\Sun-\Vec{r})},
\end{align}
and $\Vec{r}$ and $\Vec{r}_\Sun$ denote the position vectors of the orbiter and the 
Sun, respectively, measured from the center of mass of the planet (see Fig.~\ref{f:vectors}). In what follows, hats over vectors will denote directions. In particular,

\begin{equation}
\hat{\Vec{r}}=\frac{\Vec{r}}{r}, \qquad \hat{\Vec{r}}_\Sun=\frac{\Vec{r}_\Sun}{r_\Sun},
\end{equation}
with $r_\Sun=\sqrt{\Vec{r}_\Sun\cdot\Vec{r}_\Sun}$.
\par

Note that, due to the assumption of Keplerian motion, the planet-Sun distance is given by the polar equation

\begin{equation} \label{sunconic}
r_\Sun=\frac{a_\Sun(1-e_\Sun^2)}{1+e_\Sun\cos{f}_\Sun},
\end{equation}
where the semimajor axis $a_\Sun$ and the eccentricity $e_\Sun$ of the solar orbit are constant. The computation of the true anomaly $f_\Sun = {f_\Sun}(t;a_\Sun,e_\Sun)$ requires the solution of Kepler's equation \cite{Danby1992}.
\par

The contribution of the planet oblateness to the potential is given by the second-degree zonal harmonic, whose dimensionless coefficient is denoted $J_2$:

\begin{equation} \label{VJ2}
\mathcal{V}_{J_2}=\frac{\mu}{r}J_2\frac{\alpha^2}{r^2}P_2(\hat{\Vec{r}}\cdot\hat{\Vec{k}}),
\end{equation}
where $\mu$ is the gravitational parameter of the planet, $\alpha$ its equatorial radius and $\hat{\Vec{k}}$ is a unit vector in the direction of the polar axis of the primary. Hereafter, we represent a Legendre polynomial of degree $i$ as $P_i$. In particular, 

\begin{equation}
P_2=-\frac{1}{2}+\frac{3}{2} (\hat{\Vec{r}}\cdot\hat{\Vec{k}})^2.
\end{equation}
\par

For a flat plate, the acceleration due to radiation pressure is given by \cite{MilaniNobiliFarinella1987,MontenbruckGill2001}

\begin{align} \nonumber
\Vec{a}_\mathrm{srp}=& -\mathcal{P}_\mathrm{au}\frac{\mathrm{au}^2}{d_\Sun^2}\frac{A}{m}(\hat{\Vec{\iota}}_\Sun\cdot\hat{\Vec{n}}) \\
& \times\left[(1-\gamma)\hat{\Vec{\iota}}_\Sun 
+2\gamma(\hat{\Vec{\iota}}_\Sun\cdot\hat{\Vec{n}})\,\hat{\Vec{n}}\right],
\end{align}
where au denotes the astronomical unit, $\mathcal{P}_\mathrm{au}$ is the solar radiation pressure at 1\,au from the Sun, $A/m$ is the area-to-mass ratio of the object, $\hat{\Vec{n}}$ is the normal to the illuminated surface, $\hat{\Vec{\iota}}_\Sun=(\Vec{r}_\Sun-\Vec{r})/d_\Sun$ is the Sun direction from the orbiter, and the index of reflection $\gamma$ lies in the interval $(0,1)$. Recent measurements provide the value $\mathcal{P}_\mathrm{au}\approx4.5\cdot10^{-6}\,\mathrm{N/m^2}$ \cite{KoppLean2011}.
\par


While radiation pressure is a non-conservative effect, the acceleration it produces can be derived from a scalar function if we assume the plate is always facing the Sun. That is, $\hat{\Vec{\iota}}_\Sun=\hat{\Vec{n}}$, and

\begin{equation}
\Vec{a}_\mathrm{srp}=-\mathcal{P}_\mathrm{au}\frac{\mathrm{au}^2}{d_\Sun^2}\frac{A}{m}(1+\gamma)\hat{\Vec{\iota}}_\Sun.
\end{equation}
The radiation pressure acceleration can be recast as a fraction of solar gravity:

\begin{equation}
\Vec{a}_\mathrm{srp}=-\beta\mu_\Sun\frac{\Vec{r}_\Sun-\Vec{r}}{d_\Sun^3},
\end{equation}
where

\begin{equation}
\beta=\frac{\mathcal{P}_\mathrm{au}}{\mu_\Sun}\mathrm{au}^2\frac{A}{m}(1+\gamma),
\end{equation}
is the so-called lightness number, and $\mu_\Sun$ denotes the solar gravitational parameter \cite{McInnes1999}. Thus,

\begin{equation}
\Vec{a}_\mathrm{srp}=-\nabla_{\Ves{r}}\mathcal{V}_\mathrm{srp},
\end{equation}
with

\begin{equation} \label{potsrp}
\mathcal{V}_\mathrm{srp}=\beta\frac{\mu_\Sun}{d_\Sun}.
\end{equation}
\par

In our assumption both $r/d_\Sun$ and $r/r_\Sun$ are small. Therefore, the inverse of the distance $d_\Sun$ can be written as a series expansion in Legendre polynomials: 

\begin{align} \nonumber
\frac{r_\Sun}{d_\Sun}=& \; \frac{1}{\sqrt{1-2(r/r_\Sun)\hat{\Vec{r}}_\Sun\cdot\hat{\Vec{r}}+(r/r_\Sun)^2}}\\
=& \; 1+\frac{r}{r_\Sun}\hat{\Vec{r}}_\Sun\cdot\hat{\Vec{r}}+\sum_{i\ge2}\frac{r^i}{r_\Sun^i}P_i(\hat{\Vec{r}}_\Sun\cdot\hat{\Vec{r}}).
\end{align}
The constant term in the expression above does not contribute to the satellite dynamics and can be ignored. Neglecting terms $\mathcal{O}(r^2/r_\Sun^2)$ and higher, we obtain the ``potential''

\begin{equation} \label{VSRP}
\mathcal{V}_\mathrm{srp}=F_\mathrm{srp}\hat{\Vec{r}}_\Sun\cdot\Vec{r},
\end{equation}
in which

\begin{equation} \label{Fsrp}
F_\mathrm{srp}=\beta\frac{\mu_\Sun}{r_\Sun^2}
=\mathcal{P}_\mathrm{au}\frac{\mathrm{au}^2}{r_\Sun^2}\frac{A}{m}(1+\gamma)>0.
\end{equation}

Neglecting the eccentricity of the orbit of the Sun,  $r_\Sun=a_\Sun$ and the true anomaly of the Sun is replaced by its longitude:

\begin{equation} \label{sunlon}
\lambda_\Sun=n_\Sun{t},
\end{equation}
where $n_\Sun$ is the mean motion of the Sun. Therefore, the magnitude of the acceleration $F_\mathrm{srp}$ becomes constant. 

\section{Hamiltonian approach} \label{s:Hamsol}

The perturbed Keplerian motion  under the disturbing forces described by Eqs.~(\ref{VJ2}) and (\ref{VSRP}) admits a Hamiltonian formulation. The Hamiltonian must be written in terms of a set of canonical variables. A common choice for Keplerian motion is Delaunay variables $(\ell,g,h,L,G,H)$. They are usually described in terms of the standard set of Keplerian elements: semimajor axis, eccentricity, inclination, longitude of the ascending node, argument of the periapsis, and mean anomaly $(a,e,I,\Omega,\omega,M)$.

\begin{equation} \label{delaunays}
\begin{array}{cl}
\ell=& \; M, \\
g=& \; \omega, \\
h=& \; \Omega, \\
L=& \; \sqrt{\mu{a}}, \\
G=& \; L\eta, \\
H=& \;G\cos{I},
\end{array}
\end{equation}
where $\eta=\sqrt{1-e^2}$.
\par

Thus, we have a time-dependent, three-degree-of-freedom Hamiltonian

\begin{equation} \label{Hamosc}
\mathcal{H}=\mathcal{H}(\ell,g,h,L,G,H,t)\equiv\mathcal{H}_\mathrm{Kepler}+\mathcal{V}_{J_2}+\mathcal{V}_\mathrm{srp},
\end{equation}
where

\begin{equation} \label{HamKep}
\mathcal{H}_\mathrm{Kepler}=-\frac{\mu^2}{2L^2}=-\frac{1}{2}nL,
\end{equation}
is the term corresponding to the restricted two-body problem, and $n=\sqrt{\mu/a^3}=\mu^2/L^3$ is the orbiter's mean motion.
The explicit appearance of time in the Hamiltonian (\ref{Hamosc}) originates from the longitude of the Sun in Eq.~(\ref{VSRP}). Time is conveniently eliminated in a rotating frame, with angular velocity $n_\Sun$, in which the first axis is aligned with the Earth-Sun direction.

In the rotating frame the definition of the Delaunay elements is unchanged except for $h=\Omega-\lambda_\Sun$. To preserve the Hamiltonian character of the rotating frame formulation, we must include the Coriolis term

\begin{equation}
\mathcal{H}_\mathrm{Coriolis}=-n_\Sun\hat{\Vec{k}}_\Sun\cdot\Vec{G},
\end{equation}
where $\hat{\Vec{k}}_\Sun$ is the direction of the pole of the solar orbit, and

\begin{equation}
\Vec{G}=\Vec{r}\times\frac{\mathrm{d}\Vec{r}}{\mathrm{d}t}
\end{equation}
is the specific angular momentum of the orbiter. The latter is given by $\Vec{G}=G\hat{\Vec{h}}$, with $\hat{\Vec{h}}$ the unit vector in the direction of $\Vec{G}$. Therefore, the Hamiltonian (\ref{Hamosc}) becomes $\mathcal{K}=\mathcal{H}+\mathcal{H}_\mathrm{Coriolis}$, expressed as
\begin{align} \nonumber
\mathcal{K}=& -\frac{1}{2}nL
-n_\Sun{G}\hat{\Vec{k}}_\Sun\cdot\hat{\Vec{h}} \\ \label{Ham}
&+\frac{1}{2}\frac{\mu}{r}J_2\frac{\alpha^2}{r^2}\big[3(\hat{\Vec{r}}\cdot\hat{\Vec{k}})^2-1\big]
+F_\mathrm{srp}\hat{\Vec{r}}_\Sun\cdot\Vec{r}
\end{align}
in vectorial form.

Assuming the effects of $\mathcal{V}_{J_2}$, $\mathcal{V}_\mathrm{srp}$, and $\mathcal{H}_\mathrm{Coriolis}$ are small compared to $\mathcal{H}_\mathrm{Kepler}$ in Eq.~(\ref{HamKep}), say $\mathcal{O}(\epsilon)$, $\mathcal{K}$ is a perturbation Hamiltonian. Its relevant dynamical features become apparent after filtering the highest frequencies of the motion introduced by the disturbing terms.

\subsection{Short-period elimination} \label{s:shortperiod}

The elimination of the high frequencies is routinely approached with the help of perturbation methods \cite{Nayfeh2004,KahnZarmi2000,diNinoLuongo2022}. In our case, we apply canonical perturbation theory \cite{Poincare1892vII,FerrazMello2007}. More precisely, we rely on the Hamiltonian version of the method of Lie transforms \cite{Hori1966,Deprit1969,Kamel1970,Henrard1970Lie} due to its generality and versatility. It can be applied to different kinds of perturbation problems \cite{Deprit1981,PalacianVanegasYanguas2017,Lara2020,LaraMasatColombo2023}, not limited to the standard case of perturbed harmonic oscillators \cite{GiorgilliGalgani1978,Ferreretal1998ii,MarchesielloPucacco2016,Lara2017}.

Thus, we remove short-period terms by means of a canonical transformation
\begin{equation}
\mathcal{T}_\epsilon:(\ell,g,h,L,G,H;\epsilon)\mapsto(\ell',g',h',L',G',H')
\end{equation}
to prime (mean) variables, such that
\begin{equation}
\mathcal{T}_\epsilon\circ\mathcal{K}=\mathcal{K}'(-,g',h',L',G',H')+\mathcal{O}(\epsilon^2).
\end{equation}
The transformation converting $\mathcal{K}$ into
\begin{equation} \label{Knew}
\mathcal{K}'=\langle\mathcal{K}\rangle_\ell(-,g',h',L',G',H'),
\end{equation}
can be derived from the generating function
\begin{equation} \label{gen}
\mathcal{W}=\frac{1}{n}\int(\mathcal{K}-\langle\mathcal{K}\rangle_\ell)\,\mathrm{d}\ell+\mathcal{O}(\epsilon^2).
\end{equation}
More precisely, for any function of the Delaunay original variables $\xi=\xi(\ell,g,h,L,G,H)$, we can compute its transformation in terms of the prime variables $\mathcal{T}_\epsilon\circ\xi=\xi'(\ell',g',h',L',G',H')$ from

\begin{equation} \label{direct}
\xi=\xi'+\{\xi,\mathcal{W}\}.
\end{equation}
The Poisson bracket encompassing the short-period corrections must be written in prime variables for direct corrections (\ref{direct}). Conversely, $\xi'$ is written in terms of the original variables using the inverse transformation

\begin{equation} \label{inverse}
\xi'=\xi-\{\xi,\mathcal{W}\},
\end{equation}
where the Poisson bracket is evaluated in the original, non-primed variables. Obviously, this transformation is also applicable when $\xi$ is one of the Delaunay variables. Extensive details on the perturbation approach can be found in the original references \cite{Hori1966,Deprit1969}, or in modern textbooks such as \cite{BoccalettiPucacco1998v2,Lara2021}.

The short-period terms of Eq.~(\ref{Ham}) are revealed projecting the position vector $\Vec{r}$ in the apsidal frame $(\hat{\Vec{e}},\hat{\Vec{b}},\hat{\Vec{h}})$, where $\hat{\Vec{e}}=\Vec{e}/e$ is the direction of the eccentricity vector

\begin{equation}
\Vec{e}=\frac{1}{\mu}\frac{\mathrm{d}\Vec{r}}{\mathrm{d}t}\times\Vec{G}-\hat{\Vec{r}},
\end{equation}
and $\hat{\Vec{b}}=\hat{\Vec{h}}\times\hat{\Vec{e}}$. Thus,

\begin{equation}
\Vec{r}=\hat{\Vec{e}}(\Vec{r}\cdot\hat{\Vec{e}})+\hat{\Vec{b}}(\Vec{r}\cdot\hat{\Vec{b}}).
\end{equation}
Using standard relations of the ellipse:

\begin{align} \nonumber
\Vec{r}=& \big(\hat{\Vec{e}}\cos{f}+\hat{\Vec{b}}\sin{f}\big)r \\ \label{radialapsidal}
=& \big[\hat{\Vec{e}}(\cos{u}-e)+\hat{\Vec{b}}\eta\sin{u}\big]a,
\end{align}
where $f$ and $u$ denote the true and eccentric anomalies, and

\begin{equation}
r=\frac{a\eta^2}{1+e\cos{f}},
\end{equation}
from the conic equation.

In a preliminary step, the Hamiltonian (\ref{Ham}) is written as

\begin{align} \nonumber
\mathcal{K} &= -\frac{1}{2}nL
-n_\Sun{L}\eta\hat{\Vec{k}}_\Sun\cdot\hat{\Vec{h}}
+\frac{1}{3}n_*L\frac{a^2}{r^2}\frac{1}{\eta^2}\Psi(f) \\ \label{Kfu}
& +\frac{2}{3}n_\mathrm{srp}L\big[(\hat{\Vec{e}}\cdot\hat{\Vec{r}}_\Sun)(\cos{u}-e)+(\hat{\Vec{b}}\cdot\hat{\Vec{r}}_\Sun)\eta\sin{u}\big],
\end{align}
where $\mu=n^2a^3$, $n=L/a^2$, $G=L\eta$, and

\begin{align} \label{nJ2}
n_*=&\; \frac{3}{2}nJ_2\frac{\alpha^2}{a^2}, \\ \label{nsrp}
n_\mathrm{srp}=&\; \frac{3}{2}\frac{F_\mathrm{srp}}{na},
\end{align}

\begin{equation}
\Psi\equiv\big[3(\hat{\Vec{r}}\cdot\hat{\Vec{k}})^2-1\big](1+e\cos{f}).
\end{equation}
It can be shown that

\begin{align} \nonumber
\Psi= & \;
\frac{3}{2}(\hat{\Vec{b}}\cdot\hat{\Vec{k}})(\hat{\Vec{e}}\cdot\hat{\Vec{k}})(e\sin{f}+2\sin2f+e\sin3f) \\ \nonumber
& -\frac{3}{4}[(\hat{\Vec{b}}\cdot\hat{\Vec{k}})^2-(\hat{\Vec{e}}\cdot\hat{\Vec{k}})^2](e\cos3f+2\cos2f) \\ \nonumber
& +\frac{1}{4}[3(\hat{\Vec{b}}\cdot\hat{\Vec{k}})^2+9 (\hat{\Vec{e}}\cdot\hat{\Vec{k}})^2-4]e\cos{f} \\ \label{SRP:J2series}
& + \frac{1}{2}[3(\hat{\Vec{b}}\cdot\hat{\Vec{k}})^2+3(\hat{\Vec{e}}\cdot\hat{\Vec{k}})^2-2].
\end{align}

The average of the Hamiltonian (\ref{Kfu}) over the mean anomaly is obtained in closed form using the basic differential relations of Keplerian motion

\begin{equation}
\mathrm{d}\ell=\frac{r}{a}\mathrm{d}u=\left(\frac{r}{a}\right)^2\frac{1}{\eta}\mathrm{d}f.
\end{equation}
Substituting the expressions $r/a=1-e\cos{u}$ and $r/a=\eta^2/(1+e\cos{f})$, we obtain

\begin{align} \nonumber
\langle\mathcal{K}\rangle_\ell=& -\frac{L}{2}n
-n_\Sun{L}\eta\hat{\Vec{k}}_\Sun\cdot\hat{\Vec{h}}
-Ln_\mathrm{srp}\hat{\Vec{r}}_\Sun\cdot\Vec{e} \\ \label{Hamav}
&+\frac{L}{6\eta^3}n_*\big[3(\hat{\Vec{b}}\cdot\hat{\Vec{k}})^2+3(\hat{\Vec{e}}\cdot\hat{\Vec{k}})^2-2\big].
\end{align}

The generating function of the short-period elimination is computed from Eq.~(\ref{gen}):

\begin{align} \nonumber
\mathcal{W}=& \frac{L}{12\eta^3}\frac{n_*}{n}\Big\{
\big[6(\hat{\Vec{b}}\cdot\hat{\Vec{k}})^2+6(\hat{\Vec{e}}\cdot\hat{\Vec{k}})^2-4\big](f-\ell) \\ \nonumber
& +\big[3(\hat{\Vec{b}}\cdot\hat{\Vec{k}})^2 +9(\hat{\Vec{e}}\cdot\hat{\Vec{k}})^2-4\big]e\sin{f} \\ \nonumber
& -\big[(\hat{\Vec{b}}\cdot\hat{\Vec{k}})^2-(\hat{\Vec{e}}\cdot\hat{\Vec{k}})^2\big](e\sin3f+3\sin2f) \\ \nonumber
& +2(\hat{\Vec{b}}\cdot\hat{\Vec{k}})(\hat{\Vec{e}}\cdot\hat{\Vec{k}})(3e\cos{f}+3\cos2f+e\cos3f) \Big\} \\ \nonumber
& +\frac{n_\mathrm{srp}}{n}\frac{L}{6}\Big\{
 \big[2\left(2-e^2\right)\sin{u}-e\sin2u\big]\hat{\Vec{e}}\cdot\hat{\Vec{r}}_\Sun \\ \label{Wfu}
& -\eta\left(4\cos{u}-e\cos2u\right)\hat{\Vec{b}}\cdot\hat{\Vec{r}}_\Sun\Big\}+\mathcal{C},
\end{align}
where $\mathcal{C}\equiv\mathcal{C}(-,g,h,L,G,H)$ is an arbitrary function arising from the quadrature in Eq.~(\ref{gen}). While any choice of $\mathcal{C}$ would be valid from the point of view of the perturbation approach, it is common practice to select a $\mathcal{W}$ that only includes short-period terms. That is, $\langle\mathcal{W}\rangle_\ell=0$, in which case we determine $\mathcal{C}=\langle\mathcal{C}-\mathcal{W}\rangle_\ell$ form Eq.~(\ref{Wfu}). Using known primitives from the literature \cite{Kozai1962AJ,Kelly1989}, we obtain

\begin{align*}
\mathcal{C}=& 
\frac{L}{6}\frac{n_*}{n}\frac{e^2}{\eta^3}\frac{1+2\eta}{(1+\eta)^2}(\hat{\Vec{b}}\cdot\hat{\Vec{k}})(\hat{\Vec{e}}\cdot\hat{\Vec{k}}) -\frac{n_\mathrm{srp}}{n}\frac{L}{3}e\eta\hat{\Vec{b}}\cdot\hat{\Vec{r}}_\Sun.
\end{align*}
The transformations from mean to osculating variables and viceversa, given in Eqs.~(\ref{direct}) and (\ref{inverse}), are then computed evaluating Poisson brackets.

Finally, original variables are replaced with mean values in Eq.~(\ref{Hamav}) to obtain the transformed Hamiltonian $\mathcal{K}'$ (\ref{Knew}). This requires evaluating the frequencies $n_*$ and $n_\mathrm{srp}$ (Eqs.~(\ref{nJ2}) and (\ref{nsrp})) in prime variables. After neglecting higher-order terms, $\ell'$ becomes cyclic. In consequence, $L'$, $a=a(L')$, and the frequencies $n_*$
 and $n_\mathrm{srp}$ are integrals of the truncated Hamiltonian in the new variables.

\subsection{Long-term dynamics} \label{s:longterm}

The long-term dynamics can be studied after neglecting the constant Keplerian term in Eq.~(\ref{Hamav}):

\begin{align} \nonumber
\mathcal{K}'=& -L'n_\Sun\eta\hat{\Vec{k}}_\Sun\cdot\hat{\Vec{h}}
-L'n_\mathrm{srp}e\hat{\Vec{r}}_\Sun\cdot\hat{\Vec{e}} \\ \label{Hamlt}
& -\frac{L'n_*}{6\eta^3}\big[3(\hat{\Vec{h}}\cdot\hat{\Vec{k}})^2-1\big],
\end{align}
where all terms are expressed using the prime Delaunay variables, and we substituted the identity

\begin{equation}
(\hat{\Vec{e}}\cdot\hat{\Vec{k}})^2+(\hat{\Vec{b}}\cdot\hat{\Vec{k}})^2+(\hat{\Vec{h}}\cdot\hat{\Vec{k}})^2=1.
\end{equation}


The long-term dynamics are obtained from the numerical integration of the Hamilton equations. 
Denoting

\begin{equation}
\Vec\eta=\eta\hat{\Vec{h}}=\Vec{G}/L',
\end{equation}
the flow of the Hamiltonian (\ref{Hamlt}) can be written in dimensionless, non-canonical, symmetric form

\begin{align} \label{dydt}
\frac{\mathrm{d}\Vec{\eta}}{\mathrm{d}t}=& \;
n_\Sun\Vec{\eta}\times\hat{\Vec{k}}_\Sun +n_\mathrm{srp}\Vec{e}\times\hat{\Vec{r}}_\Sun
+\frac{n_*}{\eta^5}(\Vec\eta\cdot\hat{\Vec{k}})\Vec{\eta}\times\hat{\Vec{k}},
\\ \nonumber
\frac{\mathrm{d}\Vec{e}}{\mathrm{d}t}=& \;
n_\Sun\Vec{e}\times\hat{\Vec{k}}_\Sun +n_\mathrm{srp}\Vec{\eta}\times\hat{\Vec{r}}_\Sun
+\frac{n_*}{\eta^5}(\Vec\eta\cdot\hat{\Vec{k}})\Vec{e}\times\hat{\Vec{k}}
\\ \label{dedt}
&+\frac{n_*}{2\eta^5}\Big[1-\frac{5}{\eta^2}(\Vec\eta\cdot\hat{\Vec{k}})^2\Big]\Vec{e}\times\Vec\eta.
\end{align}
These differential equations are redundant due to the orthogonality of $\Vec{e}$ and $\Vec\eta$. The symmetric character of the vectorial formulation allows for an efficient implementation in software, as reported in \cite{LaraRosengrenFantino2020,SanJuanLopezLara2024}. Retaining only first-order effects, Eqs.~(\ref{dydt})--(\ref{dedt}) can be obtained adding the first terms of the mean variations of the gravitational potential to those of the problem with radiation pressure only. See Eqs.~(29)--(30) in \cite{SanJuanLopezLara2024} and Eqs.~(9.8)--(9.9) in \cite{Lara2021}.

The differential system Eqs.~(\ref{dydt})--(\ref{dedt}) approximates the averaged dynamics when the three frequencies $n_\Sun$, $n_*$, and $n_\mathrm{srp}$ are of comparable magnitude ---as required by the perturbation approach. Situations where this assumption applies have been discussed in the literature. As an example, an object with area-to-mass ratio $A/m=408\,\mathrm{cm^2/gr}$ describing an elliptical path with a semimajor axis of $17800$ km around the Earth has $n_\Sun=n_*/0.275=n_\mathrm{srp}/0.295$. See Table 2 of \cite{KrivovGetino1997} where $C\equiv{n}_\mathrm{srp}/n_\Sun$ and $W\equiv{n}_*/n_\Sun$.
\par

\section{The coplanar manifold} \label{s:coplanar}

While no closed-form integral of the Hamiltonian flow (\ref{Hamlt}) is available, there is a coplanar invariant manifold. If we neglect the axial tilt of the planet, i.e. $\hat{\Vec{k}}=\hat{\Vec{k}}_\Sun$, equatorial orbits do not experience changes in inclination. If the spacecraft starts in an equatorial orbit, the variation of $\Vec\eta$ given by Eq.~(\ref{dydt}) has the direction of $\Vec\eta$, and the motion is constrained to the plane of the equator.

We can study this particular invariant manifold by setting $\hat{\Vec{k}}_\Sun\cdot\hat{\Vec{h}}=\hat{\Vec{h}}\cdot\hat{\Vec{k}}=1$ and $\hat{\Vec{r}}_\Sun\cdot\hat{\Vec{e}}=\cos\theta$ in Eq.~(\ref{Hamlt}). The polar angle $\theta$ formed by the directions of the Sun and the orbit periapsis\footnote{Some authors use the supplementary angle of $\theta$.} is the conjugate coordinate to the specific angular momentum $\Theta=G'$. Then,

\begin{equation} \label{Hamcop}
\mathcal{K}_\mathrm{coplanar}=-L'\left(n_\Sun\eta+\frac{n_*}{3\eta^3}+n_\mathrm{srp}e\cos\theta\right).
\end{equation}
Recall that $\eta=\Theta/L'$ from Eq.~(\ref{delaunays}).

\subsection{Equilibria}

Note that the rates of change of $\theta$ and $\Theta$ 

\begin{align} \label{hameqz}
\frac{\mathrm{d}\theta}{\mathrm{d}t}= \phantom{-}\frac{\partial\mathcal{K}_\mathrm{coplanar}}{\partial\Theta}=& \;
\frac{n_*}{\eta^4}-n_\Sun+n_\mathrm{srp}\frac{\eta}{e}\cos\theta, 
\\ \label{hameqG}
\frac{\mathrm{d}\Theta}{\mathrm{d}t}= -\frac{\partial\mathcal{K}_\mathrm{coplanar}}{\partial\theta}=&
-n_\mathrm{srp}L'e\sin\theta,
\end{align}
vanish when $\theta=0$ or $\theta=\pi$, and

\begin{equation}
(n_*-n_\Sun\eta^4)e\pm{n}_\mathrm{srp}\eta^5=0,
\label{eq:condition}
\end{equation}
where the sign depends on the value of $\theta$. Equation~(\ref{eq:condition}) can be recast into

\begin{equation}
(n_*-n_\Sun\eta^4)^2e^2-{n}_\mathrm{srp}^2\eta^{10}=0,
\end{equation}
which is always valid and is a quintic polynomial in $\eta^2$. Replacing $e^2=1-\eta^2$ and expanding the factors gives

\begin{equation} \label{quintic}
(\tilde{n}_\mathrm{srp}^2+1)\eta^{10}-\eta^8-2\tilde{n}_*\eta^6+2\tilde{n}_*\eta^4+\tilde{n}_*^2\eta^2-\tilde{n}_*^2=0,
\end{equation}
in which the radiation pressure parameter

\begin{equation}
\tilde{n}_\mathrm{srp}=\frac{n_\mathrm{srp}}{n_\Sun}
\end{equation}
increases with the area-to-mass ratio. The oblateness parameter 
\begin{equation}
\tilde{n}_*=\frac{n_*}{n_\Sun}
\end{equation}
decreases when the semi-major axis grows. 
From Descartes' rule of signs, Eq.~(\ref{quintic}) has either 3 or 1 real roots, corresponding to eccentricity values for which the periapsis remains frozen. In general, the roots of Eq.~(\ref{quintic}) must be computed numerically from given values of $\tilde{n}_\mathrm{srp}$ and $\tilde{n}_*$. However, because the resultant of the quintic polynomial and its derivative with respect to $\eta$ must vanish for multiple roots, we succeeded in computing analytically the bifurcation line $\tilde{n}_\mathrm{srp}=\tilde{n}_\mathrm{srp}(\tilde{n}_*)$ that separates the regions of the parameters plane admitting one or three equilibria. We first compute the discriminant $\Delta$ of Eq.~(\ref{quintic}):

\begin{align} \nonumber 
\Delta=& 1024\big[3125 \tilde{n}_* \tilde{n}_{\text{srp}}^4
+32\tilde{n}_*(8 \tilde{n}_*^2-25 \tilde{n}_*+125) \tilde{n}_{\text{srp}}^2
\\
& +256 (\tilde{n}_*-1){}^3\big]^2\tilde{n}_*^{16}\tilde{n}_{\text{srp}}^8(\tilde{n}_{\text{srp}}^2+1).
\end{align}

Disregarding the degenerate cases $\tilde{n}_*=0$ and $\tilde{n}_\mathrm{srp}=0$, setting the discriminant to zero yields a biquadratic polynomial in $\tilde{n}_\mathrm{srp}$, whose coefficients are polynomials in $\tilde{n}_*$. Solving for $\tilde{n}_\mathrm{srp}$ gives
\begin{equation} \label{bili}
\tilde{n}_\mathrm{srp}=\frac{4\sqrt{5}}{125}\tilde{n}_*\sqrt{\frac{5-\tilde{n}_*}{\tilde{n}_*}\left[\Big(4+\frac{5}{\tilde{n}_*}\Big)^{\frac{3}{2}}-\frac{25}{\tilde{n}_*}\right]-8 }.
\end{equation}

\begin{figure}[htbp]
\center
\includegraphics[scale=0.8]{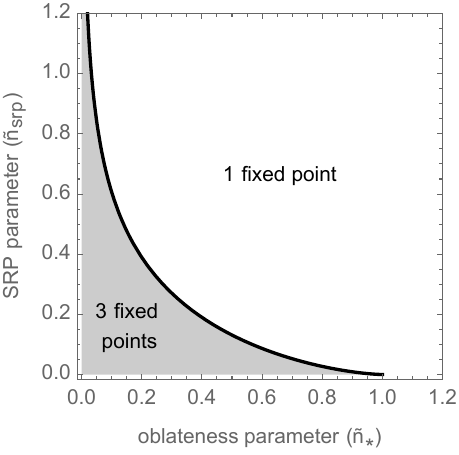}
\caption{Regions of the parameters plane with different numbers of fixed points (shaded and white areas) of the coplanar flow.
}
\label{f:planarJ2}
\end{figure}

The bifurcation line given by Eq.~(\ref{bili}) is represented in the parameters plane $\tilde{n}_*$-$\tilde{n}_\mathrm{srp}$ with a black curve in Fig.~\ref{f:planarJ2}. 
Below this line, there are always three equilibria (shaded area of Fig.~\ref{f:planarJ2}). Two of them merge at the bifurcation line and cease to exist above it (light area of Fig.~\ref{f:planarJ2}), where just one remains. We will show later that this bifurcation line is of the saddle-node type. Because the line only exists in the interval $0<\tilde{n}_*\le1$, there is only one fixed point when $n_*>n_\Sun$.

\subsection{Changes of local nature of the reduced flow}

\begin{figure*}[htbp]
\center
\includegraphics[scale=0.8]{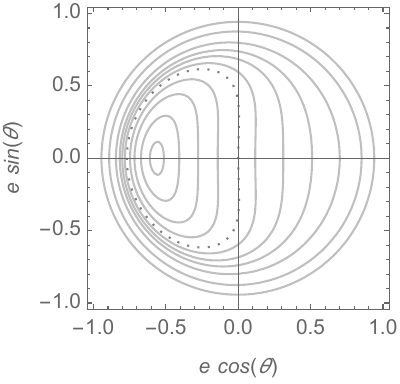}\qquad\includegraphics[scale=0.75]{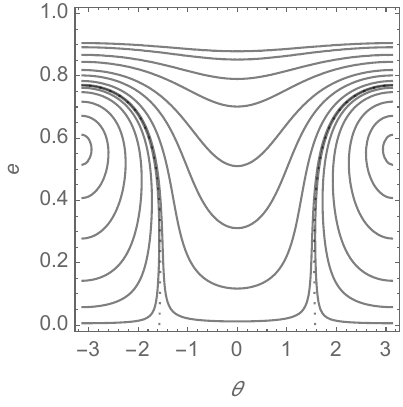} \\
\includegraphics[scale=0.8]{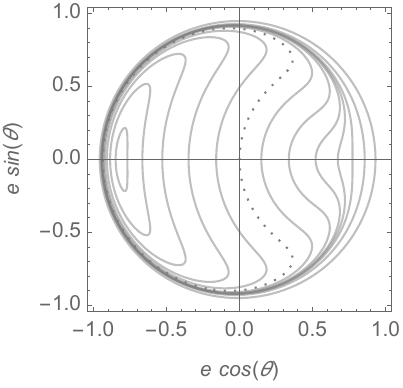}\qquad\includegraphics[scale=0.75]{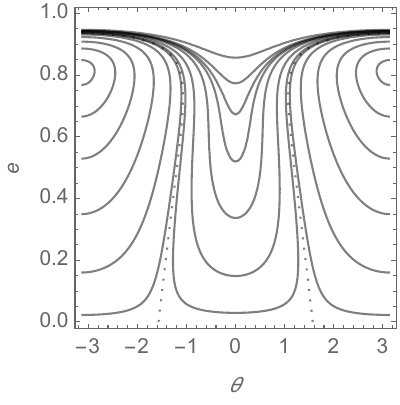} \\
\includegraphics[scale=0.8]{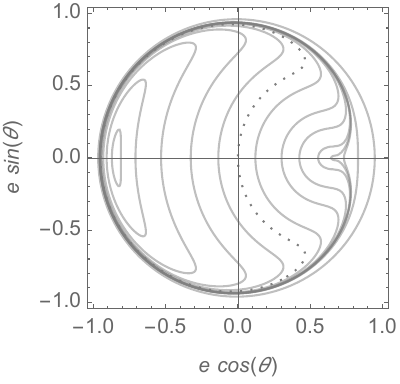}\qquad\includegraphics[scale=0.75]{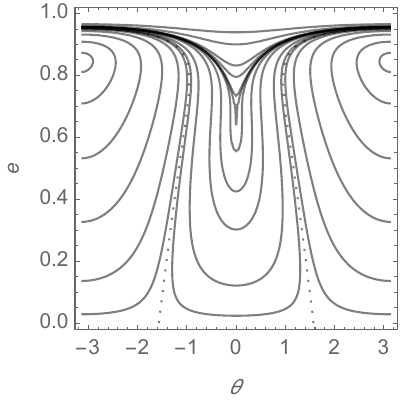}
\caption{Coplanar orbits for $\tilde{n}_\mathrm{srp}=3^{-1/2}$ and $\tilde{n}_*=0.85$ (top), $0.16$ (center) and $0.11$ (bottom). Dotted contours mark transitions between rotation and oscillation of the eccentricity vector.
}
\label{f:coupledflow3}
\end{figure*}

For a given point $(\tilde{n}_*,\tilde{n}_\mathrm{srp})$ of the parameters plane, the reduced flow can be visualized without integrating Eqs.~(\ref{hameqz})--(\ref{hameqG}) by plotting contours of the Hamiltonian (scaled by $L'n_\Sun$) in eccentricity vector diagrams:

\begin{equation} \label{HcoplanarJ2sc}
\mathcal{K}'=-\eta-\frac{\tilde{n}_*}{3\eta^3}-\tilde{n}_\mathrm{srp}e\cos\theta.
\end{equation}
The reduced flow in the region above the bifurcation line (light area of Fig.~\ref{f:planarJ2}) is shown in Fig.~\ref{f:coupledflow3} for decreasing values of $\tilde{n}_*$ and a constant value $\tilde{n}_\mathrm{srp}=3^{-1/2}$. The latter has been chosen because it represents a typical case for moderate radiation pressure perturbations, see~\cite{MignardHenon1984}. For clarity, the phase plots are depicted in both eccentricity vector representation $(e\cos\theta,e\sin\theta)$ and cylindrical map form $(\theta,e)$. Dotted contours in Fig.~\ref{f:coupledflow3} correspond to the manifold

\begin{equation} \label{K*0}
\mathcal{K}'_0=-1-\frac{1}{3}\tilde{n}_*,
\end{equation}
of orbits that become temporarily circular, obtained making $e=0$, and hence $\eta=1$ in Eq.~(\ref{HcoplanarJ2sc}). The top plots of Fig.~\ref{f:coupledflow3} illustrated a situation far from the bifurcation. We observe an interior region of orbits where the periapsis oscillates around the elliptic fixed point with $\theta=\pi$, and an exterior region with rotating periapsis. They are separated by the dotted contour of the $\mathcal{K}'_0$ manifold. As shown in the middle section of Fig.~\ref{f:coupledflow3}, the interior region of orbits with oscillating periapsis becomes larger when approaching the bifurcation line, while the flow bends towards the axis of abscissas. When the saddle-node bifurcation occurs (bottom pane of Fig.~\ref{f:coupledflow3}) a cusp appears on the symmetry axis.

\begin{figure*}[htbp]
\center
\includegraphics[scale=0.8]{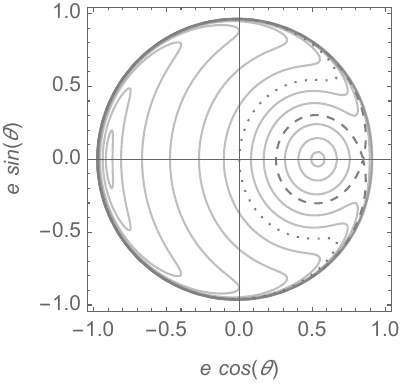} \qquad\includegraphics[scale=0.75]{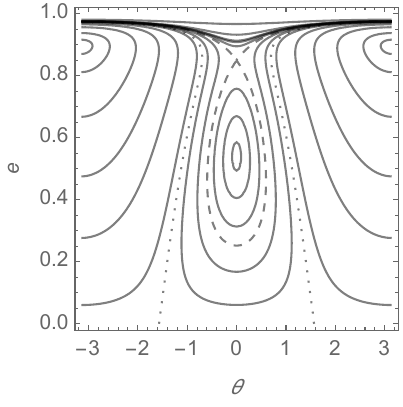} \\
\includegraphics[scale=0.8]{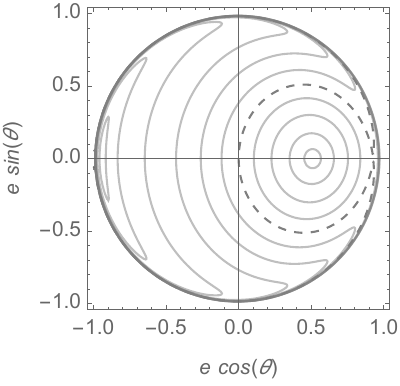}\qquad\includegraphics[scale=0.75]{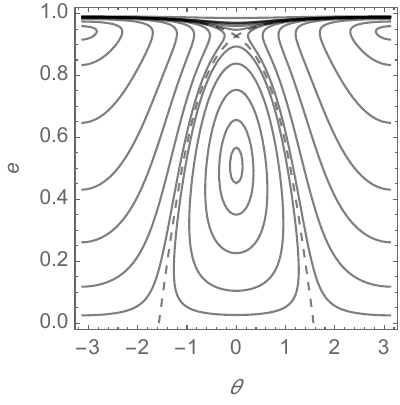} \\
\includegraphics[scale=0.8]{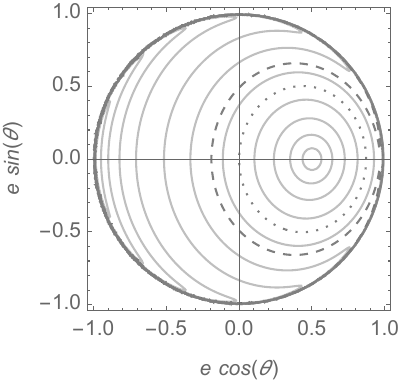}\qquad\includegraphics[scale=0.75]{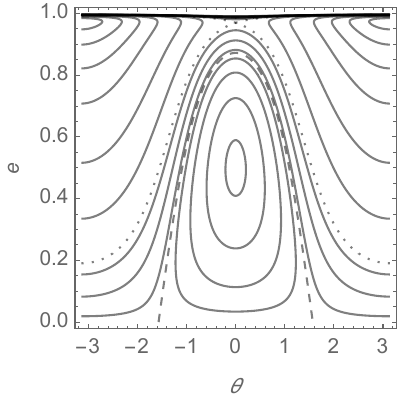}
\caption{Coplanar orbits for $\tilde{n}_\mathrm{srp}=3^{-1/2}$ and $\tilde{n}_*=0.05$ (top), $0.015$ (center), and $0.003$ (bottom). Dashed contour is the energy manifold of the hyperbolic fixed point, dotted lines ones separate regions with rotating and oscillating periapsis.
}
\label{f:coupledflow1}
\end{figure*}

Below the bifurcation line (shaded area of Fig.~\ref{f:planarJ2}) we find two additional fixed points, one elliptic and one hyperbolic, both with periapsis at $\theta=0$. The energy manifold  $\mathcal{K}'_\mathrm{U}$ of the hyperbolic fixed point plays a fundamental role in the qualitative changes experienced by the flow. As shown in the top plots of Fig.~\ref{f:coupledflow1}, an additional region of orbits with oscillating periapsis exists centered on the elliptic fixed point with $\theta=0$. It is bounded by the dashed contour corresponding to $\mathcal{K}'_\mathrm{U}$, which splits the orbits with rotating periapsis in two subsets: one between $\mathcal{K}'_0$ and $\mathcal{K}'_\mathrm{U}$, and the other, which is made of highly eccentric orbits, bounded by the exterior branch of $\mathcal{K}'_\mathrm{U}$.

\subsection{Global changes of the flow}

There are points of the parameters plane where $\mathcal{K}'_0$ may overlap to $\mathcal{K}'_\mathrm{U}$, as illustrated in the center section of Fig.~\ref{f:coupledflow1}. When this occurs, the interior region of orbits with circulating periapsis surrounding the elliptic fixed point with $\theta=\pi$ collapses to the curve defined by the interior branch of $\mathcal{K}'_\mathrm{U}$, ceasing to exist. Only three regions with different flow remain: an exterior area made of highly eccentric orbits with circulating periapsis and two interior regions with the periapsis oscillating around an elliptic fixed point.

\begin{figure}[htbp]
\center
\includegraphics[scale=0.8]{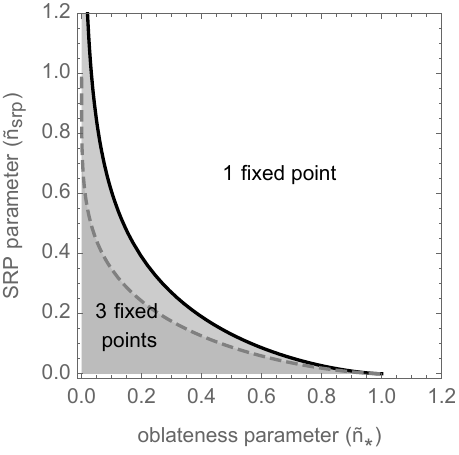}
\caption{Regions of the parameters plane of the coplanar flow with different qualitative behavior.
}
\label{f:planarJ2all}
\end{figure}

The critical condition $\mathcal{K}'_0=\mathcal{K}'_\mathrm{U}$ is expressed as a function of $\eta$ from Eqs.~(\ref{HcoplanarJ2sc}) and (\ref{K*0}):

\begin{equation}
1+\frac{1}{3}\tilde{n}_*=\tilde{n}_\mathrm{srp}e+\eta+\frac{1}{3}\frac{\tilde{n}_*}{\eta^3}.
\end{equation}
Substituting $e=\sqrt{1-\eta^2}$ yields a polynomial equation in $\eta$,

\begin{align} \nonumber
0=&\; 9(\tilde{n}_\mathrm{srp}^2+1)\eta^7
+(9\tilde{n}_\mathrm{srp}^2-9-6\tilde{n}_*)\eta^6
+\tilde{n}_*^2\eta^5
\\ \label{polyeta} &
+\tilde{n}_*^2\eta^4
+\tilde{n}_*(\tilde{n}_*+6)\eta^3 
-\tilde{n}_*^2\eta^2
-\tilde{n}_*^2\eta
-\tilde{n}_*^2,
\end{align}
which may admit either one or three real roots according to Descartes' rule. In particular, only one real root is possible if $\tilde{n}_*\le\frac{3}{2}(\tilde{n}_\mathrm{srp}^2-1)$, the condition that ensures the coefficient of $\eta^6$ in Eq.~(\ref{polyeta}) is non-negative. Therefore, the value of $\eta$ that makes $\mathcal{K}'_\mathrm{U}=\mathcal{K}'_0$ must be a common root of Eqs.~(\ref{polyeta}) and (\ref{quintic}). From the resultant of the corresponding polynomials, we obtain  the critical line $\tilde{n}_\mathrm{srp}=\tilde{n}_\mathrm{srp}(\tilde{n}_*)$ in implicit form

\begin{align} \nonumber
0=&\; 
59049\tilde{n}_{\mathrm{srp}}^8-243\tilde{n}_{\mathrm{srp}}^6(647\tilde{n}_*^2+2538\tilde{n}_*+972) \\ \nonumber
&-2\tilde{n}_{\mathrm{srp}}^4(1603\tilde{n}_*^4-13500\tilde{n}_*^3+49572\tilde{n}_*^2+291600\tilde{n}_* \\ \nonumber
& -177147)
+6\tilde{n}_{\mathrm{srp}}^2(25\tilde{n}_*^6-205\tilde{n}_*^5-6604\tilde{n}_*^4 \\ \nonumber
& -35714\tilde{n}_*^3-140967\tilde{n}_*^2-274833\tilde{n}_*-39366) \\ \label{Ke0}
& -3(\tilde{n}_*-1)^3(\tilde{n}_*+3)(\tilde{n}_*^2+14\tilde{n}_*+81)^2.
\end{align}
The solution $\tilde{n}_{\mathrm{srp}}=\tilde{n}_{\mathrm{srp}}(\tilde{n}_*)$ of Eq.~(\ref{Ke0}) is represented in Fig.~\ref{f:planarJ2all} with a dashed curve. 

After crossing the critical line a region of orbits with rotating periapsis around the elliptic fixed point with $\theta=0$ develops between the manifolds $\mathcal{K}'_0$ and $\mathcal{K}'_\mathrm{U}$ (bottom section of Fig.~\ref{f:coupledflow1}). This is in stark contrast with the situation before the crossing (top pane of the figure) where the orbits contained between these two manifolds revolve around the $\theta=\pi$ elliptic point. Thus, the line defined by Eq.~(\ref{Ke0}) marks a global transition in the flow, as opposed to the local nature of the bifurcation boundary given by Eq.~(\ref{bili}).

\subsection{The reduced flow on the sphere}

In the graphics of the previous section, the behavior of the highest eccentricity orbits is difficult to appreciate. A sphere provides better visualization \cite{Kummer1976,Deprit1983,Cushman1983,CoffeyDepritMiller1986,Lara2017}. Introducing the variables

\begin{align}
\chi_1=&\; e\eta\cos{g}, \\
\chi_2=&\; e\eta\sin{g}, \\ \label{chi3}
\chi_3=&\; \eta^2-\frac{1}{2},
\end{align}
the Hamiltonian (\ref{HcoplanarJ2sc}) can be rewritten as

\begin{equation} \label{HcoplanarJ2chi}
\mathcal{K}'=-\eta-\frac{\tilde{n}_*}{3\eta^3}-\tilde{n}_\mathrm{srp}\frac{\chi_1}{\eta}.
\end{equation}
In Eq.~(\ref{chi3}), $\eta$ can be expressed in terms of $\chi_3$: $\eta=\sqrt{\frac{1}{2}+\chi_3}$. The flow corresponds to the intersection of the manifold $\mathcal{K}'(\chi_1,-,\chi_3)=\kappa<0$ (\ref{HcoplanarJ2chi}) with the sphere

\begin{equation} \label{sphere}
\chi_1^2+\chi_2^2+\chi_3^2=\frac{1}{4},
\end{equation}
of radius $\frac{1}{2}$. Circular orbits ($e=0$, $\eta=1$) lie on the north pole of the sphere $(0,0,\frac{1}{2})$, whereas the orbits with the maximum eccentricity ($e\rightarrow1$, $\eta\rightarrow0$) collapse to the south pole $(0,0,-\frac{1}{2})$.

\begin{figure}[htbp]
\centerline{
\includegraphics[scale=0.6]{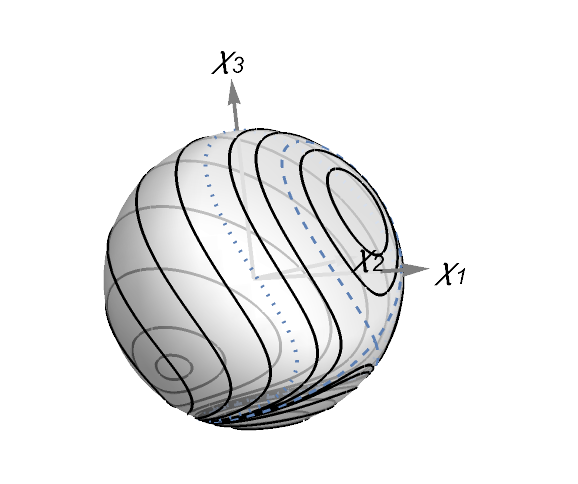} \hspace{-1.5cm}
\includegraphics[scale=0.53]{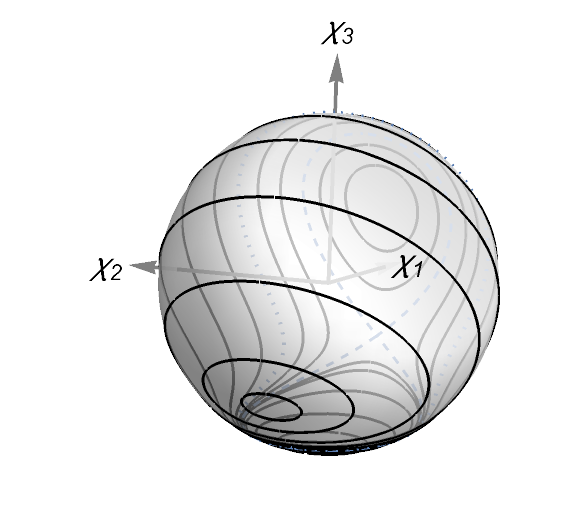}
}
\caption{Different views of the reduced coplanar flow on the sphere for $\tilde{n}_\text{srp}=3^{-1/2}$, $\tilde{n}_*=0.05$, corresponding to the first row of Fig.~\protect\ref{f:coupledflow1}.
}
\label{f:sphere}
\end{figure}

For a given energy $\kappa$, a trajectory on the sphere is computed as a sequence of points. First, we choose a value of $\chi_3$ in the interval $]-\frac{1}{2},\frac{1}{2}]$. Next, we solve $\chi_1=\chi_1(\chi_3;\kappa)$ from Eq.~(\ref{HcoplanarJ2chi}) to obtain

\begin{equation}
\chi_1=-\frac{1}{\tilde{n}_\mathrm{srp}}\left(\eta^2+\kappa\eta+\frac{1}{3}\frac{\tilde{n}_*}{\eta^2}\right),
\end{equation}
where $\eta\equiv\eta(\chi_3)$. Finally, we compute $\chi_2$ from Eq.~(\ref{sphere}):
\[
\chi_2= \pm\left(\frac{1}{4}-\chi_1^2-\chi_3^2\right)^{\!\frac{1}{2}}.
\]
The case $\tilde{n}_\text{srp}=3^{-1/2}$, $\tilde{n}_*=0.05$, previously presented in the first row of Fig.~\ref{f:coupledflow1}, is shown in Fig.~\ref{f:sphere}. It highlights the circulation of highly elliptic orbits around the south pole of the sphere. 

\section{Conclusions}

The long term behavior of a high area-to-mass ratio object orbiting a planet may undergo important qualitative changes induced by the oblateness perturbation of the central body. Moving beyond resonant cases, extensively discussed in the literature due to their interest for astrodynamics applications, we focused on coplanar orbits. Neglecting the axial tilt of the planet, the equatorial orbits of this kind of objects constitute an invariant manifold of the oblate-solar radiation pressure problem, integrable up to higher-order effects. The invariance of the coplanar manifold is evidenced by the vectorial formulation presented. The generality of the approach, based on the fundamental vectors defining the apsidal frame, and free from singularities, may offer advantages for semi-analytical integration.

Particular cases of the solution presented in this contribution can be found in the literature. We generalised the treatment using a rigorous and formal approach. We derived the generating function of the transformation for the averaging explicitly. We selected the arbitrary function on which 
the mean-to-osculating transformation  depends to ensure the latter is purely periodic. This is a prerequisite for extending the theory to second order.
We extended the existing literature by thoroughly exploring the qualitative behavior of the coplanar manifold in a two-parameter plane. One of the parameters is related to the physical characteristics of the orbiter, while the other is associated with the dynamical characteristics of the orbit. We computed analytically two critical boundaries delimiting three regions with different qualitative flow. The characteristics of the flow depend on the balance between the solar radiation pressure and the oblateness perturbations.
Local changes to the flow appear in the form of bifurcations of orbits with the periapsis frozen in the radiation pressure direction. Global variations in the flow are possible. Orbits with rotating periapsis can switch from circling the fixed point with periapsis towards the Sun, to revolving around the opposite frozen orbit.

\begin{acknowledgements}
The authors acknowledge Khalifa University of Science and Technology's internal grant CIRA-2021-65 (8474000413). ML also acknowledges partial support from the Spanish State Research Agency and the European Regional Development Fund (Projects PID2020-112576GB-C22 and PID2021-123219OB-I00, AEI/ERDF, EU). EF has been partially supported by Spanish MINECO's funds PID2020-112576GB-C21 and PID2021-1239 68NB-I00. RF received support from MINECO ``Severo Ochoa Programme for Centres of Excellence in R\&D'' (CEX2018-000797-S).
\end{acknowledgements}

\section*{Conflict of interest}
The authors declare that they have no conflict of interest.

\section*{Authors contributions}
The study conception and design were performed by M.~Lara who wrote the first draft of the manuscript. All authors collaborated on improving subsequent versions of the paper, and read and approved the final manuscript.

\section*{Funding}
The financial support for the execution of the research here presented was provided by the following grants: Khalifa University of Science and Technology's CIRA-2021-65 / 8474000413 (recipient: E. Fantino),  Spanish National funds PID2020-112576GB-C22 (recipient: M. Lara), PID2021-123219OB-I00 (recipient: M. Lara), PID2020-112576GB-C21 (recipient: E. Fantino), PID2021-123968NB-I00  (recipient: E. Fantino) and CEX2018-000797-S (recipient: R. Flores).

\section*{Data availability}
No datasets, other than the numerical results presented in the figures, have been generated as part of this research.
The information provided in this manuscript is sufficient to reproduce those results.

%
%

\bibliographystyle{spmpsci}   
\bibliography{referencesNODY_V06}   
\end{document}